\documentclass[11pt]{article}
\usepackage[a4paper,left=3cm,right=2cm,top=2.5cm,bottom=2.5cm]{geometry}
\usepackage{cite}
\usepackage[pdftex]{graphicx}
\usepackage[cmex10]{amsmath}
\usepackage{amssymb}
\usepackage{url}

\usepackage{times}
\usepackage{array}
\usepackage[tight,footnotesize]{subfigure}
\usepackage[font=footnotesize]{subfig}
\usepackage{graphicx}
\usepackage{mathtools}
\usepackage{amsmath}
\usepackage{enumitem}

\usepackage{alltt}
\usepackage{enumitem}

\def\boxx{{\vcenter{\vbox{\hrule height.3pt
          \hbox{\vrule width.3pt height6pt
          \kern6pt\vrule width.3pt}\hrule height.3pt}}\;}}

\def\impos{{\;\vcenter{\hbox{\rule[0.15mm]{1.8mm}{1.7mm}}} \;}}
\newcommand{\imposition}[1]{\xrightarrow{~~#1~~~} \kern-10pt \impos}

\def\lrarrow{\leftrightarrow \kern-8pt \rightarrow}

\def\2{\frac{1}{2}}

\newcommand{\scopepromise}[2]{\xrightarrow[#2]{#1}}

\def\beq{\begin{eqnarray}}
\def\eeq{\end{eqnarray}}
\def\bex{\begin{alltt}\small}
\def\eex{\normalsize\end{alltt}}

\newtheorem{textpromise}{Promise}[section]

\title{A Promise Theoretic Account of the Boeing 737 Max MCAS Algorithm Affair}

\author{
Jan A.\ Bergstra\\
{\small{Minstroom Research BV Utrecht, The Netherlands}}\thanks{\texttt{janaldertb@gmail.com,info@minstroomresearch.nl}}
\\~\\
Mark Burgess\\
{\small{ChiTek-i, Oslo, Norway}\thanks{\texttt{mark.burgess.oslo.mb@gmail.com}}}}

\date{23 December 2019}
\begin{document}

\maketitle

\begin{abstract}
  Many public controversies involve the assessment of statements about
  which we have imperfect information. Without a structured approach,
  it is quite difficult to develop an approach to reasoning which is
  not based on ad hoc choices.  Forms of logic have been used in the
  past to try to bring such clarity, but these fail for a variety of
  reasons.  We demonstrate a simple approach to bringing a
  standardized approach to semantics, in uncertain discourse, using
  Promise Theory. As a case, we use Promise Theory (PT) to collect and
  structure publicly available information about the case of the MCAS
  software component for the Boeing 737 Max flight control system.
\end{abstract}
\tableofcontents

\section{Introduction}

The tragic crashes of two Boeing 737 Max aircraft led to the grounding
of the worldwide fleet and an extensive investigation into the causes. The flight
safety of the aircraft design, including its software, has been called
into question, centering on a software component known as the
Maneuvering Characteristics Augmentation System (MCAS).  

In this paper, we present some initial considerations concerning the
discussion surrounding the ways in which faults and flaws may have entered the
aircraft's human-machine system\footnote{Faults refer to components
  that are unable to keep their promises, while flaws are promises
  that were inappropriately incommensurate with their intended
  goals\cite{treatise2}.}.  Software engineering safety, in
particular, is a subject in rapid flux, given the pace of software
development. It began, in earnest, with the informal definition of
failure modes \cite{ieeefail}, and was extended principally through
discussions of specific technologies and their security properties.
More recently, methodology concerning `post-mortem analysis' of
incidents and systemic learning has been developed
\cite{allspaw2,allspaw1,reed1}.  More generally, system safety---with
an emphasis on flight safety---has a long history and an extensive
literature (see for example \cite{reason,dorner, hollnagel1,dekker}),
and has been related to the wider concept of systematic
stability\cite{certainty,treatise1,treatise2}, but the tools of
analysis are largely informal and heuristic, so there is considerable
room for more constrained languages for analytic reasoning.
Introducing some aspects of Promise Theory as a tool for such analysis
is one goal of this work.

Software plays a role in nearly all complex systems today.
Increasingly, the focus rests on `algorithms'---where the term
`algorithm' enshrines a bundle of design decisions on how a system is
supposed to react under certain conditions.  An algorithm is (perhaps
boldly and certainly informally) said to `determine' the decisions
made by the system in different circumstances; but algorithms are
logical trees of possible pathways---they may also make use of data
gathered in real time or from experience, by learning techniques. 
We will return to the notion of an algorithm in the concluding section.  

The suggestion of determinism therefore overstates the capabilities of
algorithms. All we can really say is that their promises influence 
outcomes in some fashion. The more recent discussions
concerning the use of machine learning, say for self-driving
vehicles\cite{selfdrive1,selfdrive2}, illustrates how there is often a
mismatch of complexities in reliance in software algorithms.  When
notable failures occur, certain system components, including software
components, may become the subject of intense public debates,
conducted at a high level of abstraction, and thus far removed from
technical realities.

In this work, we base our analysis on the following assumptions:

\begin{itemize}
\item Algorithms, and the software components realizing them, are
  subjects for public deliberation, reflection, and scrutiny. We refer
  to this as the external assessment of algorithms and software
  components.

\item The language and notations of computer science, and of
  engineering, are often too technical and too detailed for use by the
  public at large, i.e. for informal external assessments. Moreover,
  the strict forms of logic are unhelpful in analyzing problems of
  non-trivial complexity. Some middle ground is helpful.

\item Promise Theory is a useful tool for the assessment of algorithms
  and software components, because it offers a semi-formal approach
  based on a clear model of interaction, with a decade of experience
  and application\cite{treatise1,treatise2}. It's our goal to pursue
  this avenue through the Boeing case study.
\end{itemize}
Promise Theory was originally conceived as an approach to modelling in
agent-based systems, i.e. in systems composed from independent
components, as networks, from the bottom up. PT exposes the role of
agent interactions, both semantically and dynamically. Although
originally applied in the context of distributed computer systems, the
method is general and has since been applied at larger scales and even
social and political contexts. It includes the possibility of interactions
between human and non-human agents, and the effects of the fidelity
with which agents are able to keep promises that may be offered at
varying levels of precision\cite{BergstraB2014}.

Promise Theory breaks the world broadly into agents, promises, and
assessments.  An agent may be any separable entity, i.e. one that has
the capacity to inject independent causal information to the system. A
promise is a statement of intent, and can be made by single agents or
by groups of agents (superagents). Promises typically fall into two
types: behaviours offered (sometimes called + promises) and behaviours
accepted (sometimes called - promises). The description of what
constitutes a promise may be specified in any language, in principle,
but the more precise an offer is made, the easier it is for it to be
accepted and assessed.  An assessment, on the other hand, is a
statement of the belief made by an agent about whether a promise (made by
any other agent, or itself) has been kept or not. By breaking a system
down into these parts, PT offers a surprisingly clear picture of
system completeness, which is not dogged the excessive constraints of 
formal logics.

In this work we try to find an appropriate scale and language for describing the
human-machine collaboration involved in a flight system: promises are
used to inform agents at all levels: machine to machine, machine to
human, human to machine, and system to external observers, who may be
either technical or non-technical.  We focus on the catastrophic
losses of two Boeing 737 Max flights as a case study, and we develop a
promise theoretic analysis suitable for an external assessment of the
MCAS component of Boeing 737 Max flight control software.  External
assessments may range from design choices to implementation artifacts,
and their roles during flights. This should include the concept of
ongoing development, including updates to components, which is common
in software engineering in particular.
 
\subsection{Promise Theory and modes of application}

Promise Theory can be used in a variety of ways. In this paper we use
it as a tool for structuring information, in the public discussion,
which has mainly concerned the requirements of certain software
components.  Lacking the undisclosed `inside information' of the
investigation, as most commentators do, we must be careful not to make
unsubstantiated claims. PT is helpful, nevertheless, in sorting out
claims and context where varied assessments comes into play.

The first hint that PT can be useful is the following inequality in
treatment of systems.  The term `system requirements' is most commonly
used in engineering methodology when complicated machinery and human
systems are discussed and designed, In other words, requirements and
expectations are imposed rather than stating what tolerances the
resulting implementations are able to deliver. This focus on
requirement is supposed to ensure their fitness for purpose. By
contrast, simpler off-the-shelf components are described by their
tolerances (effectively stating the limits of what they promise,
rather than what an independent party requires of them).

According to Promise Theory, requirements have the status of
`impositions'---i.e. an external pressure exerted on a system or its
builders to accept and implement certain directives and to promises
commensurate with them in response. Multiple requirements, i.e.
multiple impositions can be inconsistent and lead to uncertainty, so
it's always better to flip the discussion around to what promises
components can make independently, and then study the interactions of
those promises\cite{BergstraB2014}.  That will be our strategy here.
Regardless of the process used to form expectations about the
behaviour of system components, we choose to draw attention to what
they are able to promise, rather than what others may seek to require
of them.

In order to illustrate and clarify the available degrees of freedom
when working with PT, we distinguish five modes of application:
\begin{description}
\item [Political strategy level (PAL).] PT used to express long
  term policies, independent of individual action or preference. Here
  promisers and promisees are humans.
\item [Tactical political level (PAL).] PT used to discuss the
  behaviour of individuals and groups in elation to specific
  objectives. Here promisers and promisees are humans and social
  groups.
\item [Software requirements assessment level (SRAL)] PT used to
  discuss and reflect on the role that software components play in
  a specific context, for instance whether or not a specific software
  component, say in control of the aircraft, allows enough 
  Meaningful Human Control (MHC), see for
  instance~\cite{SantoniH2018,ElandsEtAl2019}
  and~\cite{SlijperBKB2019}) over the actions of the total system.

\item [Technical interaction theory level (PAL)] PT used as a
  framework for the description and development of theoretical
  accounts of multi-agent systems (e.g.~\cite{BergstraB2019}).
\item [Software technology level (PAL).] PT is used to specify the
  intended behaviour of software components\footnote{Promise Theory
    emerged from the study of stability and formal correctness of
    system states in computer installations as a deviation from the
    over-constraints of logical reasoning towards network processes
    \cite{burgessC1,burgessC2,burgessC4,burgesstheory}.}.
\end{description}
Generally speaking, promises explain the envelope of system behaviour,
in the form of pre-specified outcomes and language---applied to the
many software components and combinations inside systems. The great
bonus of using PT is that it can easily unify the roles of different
kinds of agent---both human and non-human---in a single framework.

\section{A litany of explicit and supposed promises by agents involved}

We begin without further delay to describe some of the promises
made by the Boeing aircraft system and its manufacturer.
The key promises set the scene, principally from the perspective of
each {\em promiser}. Here they form the basis on which to asses the
MCAS software component. Each promise has the following structure:
\beq \text{\bf Promiser Agent}~~
\scopepromise{\text{`body'}}{\text{\bf Scope}}~~ \text{\bf Promisee
  Agent} \eeq where the body of the promise represents an explication
of the intended outcome, and the scope is a list of agents who are
privy to the promise between the promiser and promisee.

\section{The main agents referred to}

 Agents will be written in bold face, and include:
\begin{itemize}
\item {\bf Boeing management} (Boeing)
\item {\bf Airline management}
\item {\bf Pilots}
\item {\bf FAA}, i.e. the Federal Aviation Authority.
\item {\bf Authors}, i.e. us.
\item {\bf Public}, i.e. the audience for public discourse.
\item {\bf Ralph Nader}, political activist and consumer advocate.
\item {\bf W. Bradley Wendel} Associate Dean for Academic Affairs and Professor of Law,
Cornell University.
\item {\bf Peter Ladkin} Professor of Computer Networks and Distributed Systems in the Faculty of Technology at the University of Bielefeld.
\item {\bf Benno Baksteen} Captain and former president of the Dutch Airline Pilots' Association.
\item {\bf DO178c} Software Engineering standards.
\end{itemize}

\begin{figure}[ht]
\begin{center}
\includegraphics[width=10cm]{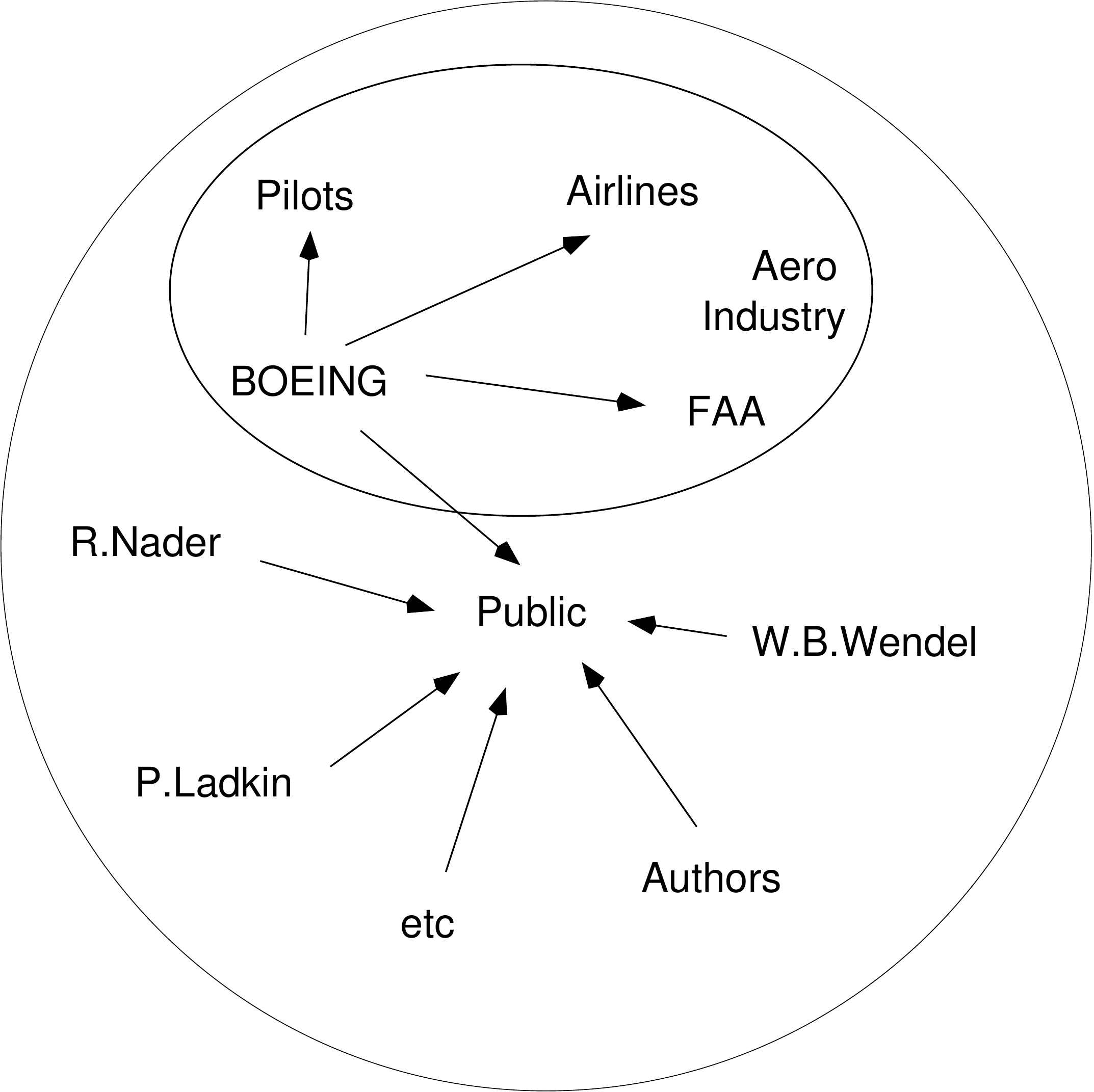}
\caption{\small The promises directed between agents. The public is a superagent containing
all the others. Most agents direct their promises publicly, but a few a solely kept within
the aerospace industry. There is a non-transparent boundary to which the general public
is not privy.}
\end{center}
\end{figure}

\subsection{Boeing as the promiser}

Three promises from Boeing's side communicate an awareness of what is
at stake. The most central and stark promise was that of continuity
in system behaviour between the longstanding 737 and the revitalized
737 Max. 

\begin{textpromise} (Model continuity promise) 

\bigskip\noindent
From {\bf Boeing management} (promiser) to {\bf Airline management} (promisee) with {\bf Pilots} in scope: 

\bigskip\noindent {\sc Promise Body}:
\begin{enumerate}[label=(\roman*)]
\item For a pilot used to a B737 NG, the B737 Max flies like B737 NG, 

\item No face to face retraining courses are needed for {\bf Pilots} upgrading to the Max,

\item No simulator training is needed for {\bf Pilot} conversion either, 

\item {\bf Pilots} can fly B737 NG and B737 Max (interchangeably), 

\item Certification of B737 Max is done, as it is the same type of B737 NG.
\end{enumerate}
{\sc End promise}
\end{textpromise}

\begin{textpromise} (MCAS hidden existence promise)

\bigskip\noindent
From {\bf Boeing} to {\bf FAA} (with no Airlines or pilots in scope, only selected Boeing engineers and FAA specialists in scope): 

\bigskip\noindent {\sc Promise Body}:
\begin{enumerate}[label=(\roman*)]
\item The MCAS software component (new in the B737 Max) takes care of
  aerodynamic differences in behaviour between B737 NG and B737 Max by
  giving additional stabilizer trim commands under certain conditions,

\item MCAS operates in such a manner that a {\bf Pilot} need not even
  know about the existence of MCAS, and therefore need not know how it
  works either.

\item Even if it is detected that the Angle of Attack (AoA) sensors disagree, that
  information need not be communicated as a warning to the pilots
  (although this warning signal can be bought at additional costs).
  For {\bf Pilots} who do not know of the existence of MCAS such information
  is of no use, and increases the risk of information overload in the
  cockpit.
\end{enumerate}
{\sc End promise}
\end{textpromise}

Item (iii) above is confirmed in~\cite{Wendel2019} where it is
considered a plausible design decision, and is mentioned
in~\cite{HuttonR2019} where it is considered an implausible design
decision.

\begin{textpromise} (Non-antistall promise) 
\label{Nas-prom}

\bigskip\noindent
{\bf Boeing} to {\bf FAA} (no-one else in scope): 

\bigskip\noindent {\sc Promise Body}:
\begin{enumerate}[label=(\roman*)]
\item MCAS is merely installed in order to have the same flight experience for pilots, 

\item MCAS is not meant to prevent stalling or any other potential
  calamitous emergency which has become more probable in the B737 Max
  compared with the B737 NG because of aerodynamic differences (which
  result from different size and placement of the engines).

\item MCAS is based on existing technology that is properly working in
  the Boeing KC-46 Pegasus tanker plane, which is a further
  development of a Boeing 767 cc).
\end{enumerate}
{\sc End promise}
\end{textpromise}
We might note that such a promise may immediately be suspected as a
deception, since promising that the addition of a software component
meant to counter the effects of mechanical changes would have no
effect on stalling is implausible in and of itself.

\subsection{Authors as promisers}

The following are some promises that we, the authors, provide for the
benefit of the reader (you), for context and reference.  The promises
in this section are of general methodological interest as a reasoning
tool.  They point to one advantage of PT over a precise {\em logical}
argument, and over free rhetoric: without PT, when stating the promise
bodies as factual information, there is a need to find evidence in
support of the assertions. Logical statements should be considered
true or false---perhaps with some probability.  

Remarkably, with promises, one can do without providing evidence, as
long as the claims are not beyond reason, because the implicit
semantics are sufficient for the purpose of establishing a the
positions semantically. This is in contrast to logical approaches,
where actual evidential inputs are needed to establish branch points
in an argument, else one faces an exponential growth of brittle and
mutually exclusive possibilities. With PT, on the other hand, a reader
who disagrees with an agent's promises (e.g. is able to find a
convincing rationale for the existence of a software component MCAS-p
in the Boeing KC-46 Pegasus) could simply decrease their trust in that
agent---as you, the reader, might decide that we---the authors of this
document---are untrustworthy, if we should make promises that violate
your perceived trust.  Indeed, this is generally true---we may have
failed to find crucial evidence, which we accept, but that is not a
reason to abstain from trying to make sense of what we can discover.
Others may offer flawed evidence, which could lead to more trust than
is warranted.


\begin{textpromise} 
\label{ETproblem}
(Existing technology questionability promise.) 

\bigskip\noindent
{\bf Authors} to {\bf Public} readers: 

\bigskip\noindent {\sc Promise Body}:
\begin{enumerate}[label=(\roman*)]
\item By MCAS-p, we refer to the MCAS-like software component developed for the Boeing KC-46 Pegasus,

\item There is no information easily and publicly available which
  explains what the role of MCAS-p is or is intended to be in the
  Boeing KC-46 Pegasus flight control system.

\item It is implausible that MCAS-p only serves to create a look and
  feel for pilots which conforms to a type which they have been flying
  before (as mentioned above).

\item It is hard to find out (and we did not succeed in determining)
  which objectives of MCAS-p have been preserved as the objectives of
  MCAS, which objectives were dropped, and which new objectives were
  assigned to MCAS w.r.t. MCAS-p.
\end{enumerate}
{\sc End promise}
\end{textpromise}

\begin{textpromise}

\bigskip\noindent {\bf Authors} to {\bf Public} readers (i.e. {\bf Public} in scope). 

\bigskip\noindent {\sc Promise Body}:
\begin{quote}
  The B737 Max problem may be considered a software problem, if only
  because its solution (according to Boeing) consists of an upgrade of
  the MCAS software component.
\end{quote}
{\sc End promise}
\end{textpromise}

\begin{textpromise}\label{featint}

  \bigskip\noindent {\bf Authors} to {\bf Public} readers (i.e. {\bf
    Public} in scope). 

\bigskip\noindent {\sc Promise Body}:
\begin{quote}
  The B737 Max software problem may be considered an instance of
  feature interaction, i.e. promises that are conditional on one
  another (See e.g.~\cite{KimblerB1989} for the concept of feature
  interactions.)
\end{quote}
{\sc End promise}
\end{textpromise}
To justify promise~\ref{featint} we mention that the following
features interacted during the calamitous events:
\begin{enumerate}
\item Repeated automatic stabilizer downwards trimming upon detection
  of excessive angle of attack, and independently of pilot's attempts
  to override the stabilizer trimming commands.
    This situation was described in detail w.r.t. the Lion Air crash in
    \cite{lionair} (a key feature introduced after the Airbus
    earlier 330 crash over the Atlantic, where it became clear that
    pilot's failed to believe that the aircraft was at risk to enter a
    stall. This suggestion has been put forward in detail
    in~\cite{Wendel2019}.).

\item The immediate predecessor aircraft model (737 NG) allows {\bf
    Pilots} to counteract runaway trim\footnote{`Runaway trim occurs
    when the Trimmable Horizontal Stabilizer (THS) or other trim
    device on the aircraft tail fails to stop at the desired position
    and continues to deflect up or down. Runaway trim can have several
    causes, including but not limited to a bad switch, a short
    circuit, or a software failure.'\cite{trim}} via yoke handling (although the
  flight manual instructs not to do so, and instead proposes (i) apply
  the manual switch off of the electric motors which move the trim,
  and (ii) thereafter use manual control until safe landing, (while
  $737$ Max 8 technically only allows the second solution of a trim
  runaway).

  (The $737$ Max design made it a feature that the instructions
  in the flight manual were the unique and only way to solve the
  problem. Thereby a parasitic feature, inherited from the 737 NG, was
  overridden and the need for training to undo habits that contradicted
  flight manual prescriptions went undetected. Recent information
  suggests that not following flight manuals is still standard
  practice to date. See~\cite{Wendel2019} for more detail.) 

\item MCAS is triggered (to change the stabilizer position by several
  degrees, in order to make the aircraft turn in a downward direction)
  by a single AOA sensor, even though a second AOA sensor produces
  highly deviating values, and even if the system has observed that
  AOA measurements have become unreliable\footnote{Readers may note
    that a promise theoretic analysis of the flight system, at the
    deeper software level, would have revealed this as an immediate
    red flag, when two agents promise data, yet only one agent's data
    was accepted. A simple count of + and - promises would have
    revealed the design flaw.}.

  (This might be arguably ok in terms of probability calculus, and
  even preferable to voting with two AOA sensors, as long as the risk
  of trimming the stabilizer when that is not required was negligible,
  that is when trigger by a faulty AOA sensor is manageable. However,
  it exposed a critical failure path.)

\item If runaway trim is detected too late and the move towards fully
  manual flying has been made, it may be too heavy going for the
  pilots to use the manual stabilizer control in order to counteract
  the trim runaway.

  (Acquiring effective manual control is not so easy as might be
  expected. Those pilots who have had hands on experience with manual
  control in the 737 NG under demanding circumstances found
  that effectuating manual control became slightly harder in the
  737 Max. Indeed manual stabilizer controls now have a slightly
  smaller radius than in the predecessor model, a feature which saves
  cockpit space, but which may require more human power and which for
  that reason may be problematic at higher speeds. High speed is
  likely to occur after a stabilizer runaway has been detected and
  reacted upon late.)

\item Hiding from the pilots the information that both AOA sensors
  disagree (a feature intended to avoid cognitive overload, in a
  critical phase, was justified by the observation that the AOA is not a
  quantity on which the human control of a commercial airliner is
  based nowadays).

  (In the 737 Max case, knowing that AOA sensors diverge in their
  readings, and that AOA measurement is unreliable for that reason is
  of no help whatsoever for the {\bf Pilots}, unless the {\bf Pilots}
  know that this may negatively impact on MCAS behaviour. Therefore
  suppressing this information was deemed a reasonable feature in the light of
  information overload prevention.)

\item The role of simulators has been complex. Simulator tests during development
missed the scenario of the Lion Air accident, which upon retrospectively
analyzing the accident could be replicated and led to the conclusion that
repeated MCAS interventions in addition to other warnings indirectly caused by
AOA disagreement increased the workload for the {\bf Pilots} too much
(see \cite{flightglob1}).
Simulator tests after redesign of the B737 Max has shown anew complications which too months to
resolve (see \cite{busins1}).
B737 Max {\bf Pilots} had only been using B737 NG simulators for training which don't
contain the MCAS software, or an appropriate abstraction of it. B737 Max simulators exist, but without MCAS simulation.\footnote{%
CockpitMax.com advertises a B737 Max simulator which does not mention
MCAS as one of the simulated flight control software components.}

\end{enumerate}

\begin{textpromise} (Contemplating the hypothetical B737 Max-minus)

\bigskip\noindent
{\bf Authors} to {\bf Public} readers ({\bf Public} in scope). 

\bigskip\noindent {\sc Promise Body}:
\begin{enumerate}[label=(\roman*)]
\item By a hypothetical B737 Max-minus, let us imagine an aircraft B737 Max without MCAS,
  i.e the airplane B737 Max modified in such a manner that ``AOA too
  high'' alerts are not processed, so that no interventions from MCAS
  may occur.

\item From the publicly available literature one cannot infer whether
  or not B737 Max-minus can be certified as a new type of aircraft.
  However, promise~\ref{Nas-prom} suggests that this would indeed be
  possible.

\item It is an open theoretical problem whether or not a software
  component MCASb exists (in a mathematical sense, i.e. can be
  designed) which transforms B737 Max-minus, which is referred to as
  the unaugmented B737 Max \cite{wiki1}
  by including MCASb, to an aircraft B737 Max (MCASb) which can be
  successfully certified as an airplane of the same type as B737 NG.
  This problem is decidable, in principle, by assuming that an airframe
  will be used not more that a predetermined maximum of hours, and
  that it suffices to recompute all desired outputs, say every millisecond.

\item The principal conceptual problem for the design of MCASb is
  whether or not it allows meaningful human (i.e. pilot) control (MHC)
  in all stages of the flight.
\end{enumerate}
{\sc End promise}
\end{textpromise}

\begin{textpromise} \label{not-agree}\label{RationalUse} 
(Rationale for the use of a single AOA sensor)

\bigskip\noindent
 {\bf Authors} to {\bf Public} readers (public in scope):

\bigskip\noindent {\sc Promise Body}:
\begin{enumerate}[label=(\roman*)]
\item If it turns out to be the case that MCAS-p (see
  promise~\ref{ETproblem}) has been designed as an anti-stall system,
  then the use of input from a single AOA sensor is only plausible (in
  hindsight).

\item If MCAS results from a natural evolution of predecessor MCAS-p,
  then the use of a single AOA sensor output is plausible in
  hindsight (while at the same time it indicates a mismatch between
  high level requirement of making the B737 Max fly like a B737
  NG, and low level requirements i.e. inheriting anti-stall functionality
  from MCAS-p.
\end{enumerate}
{\sc End promise}
\end{textpromise} 

This concludes the assertions that we infer from literature surrounding the
incident. Readers should assess each of these promised statements in the
light of their possibly greater access to information.

\subsection{Other external observers as promisers}

Finally, conducive to the argument of software flaws, let us consider some more
points.

\begin{textpromise} \label{FalseAlarm}(False alarm.) 

\bigskip\noindent
{\bf Benno Baksteen} to {\bf Public}: 

\bigskip\noindent {\sc Promise Body}:
\begin{enumerate}[label=(\roman*)]
\item There is  no problem with B737 Max at this moment (March 13 2019)\cite{benno1}.
\item The step to ground B737 Max aircraft in China is overly cautious
  and is not well-founded in facts.

\end{enumerate}
{\sc End promise}
\end{textpromise}

Our assessment: this promise turned out to be impossible to keep
upright, and thereby sets and end to the trust we have in Benno
Baksteen as an aviation safety expert.

\begin{textpromise}(MCAS is a patch which provides stability and anti stall protection). 

\bigskip\noindent
{\bf Ralph Nader} to {\bf Public} (all {\bf Public} in scope):

\bigskip\noindent {\sc Promise Body}:
\begin{enumerate}[label=(\roman*)]
\item MCAS is a software patch which compensates for aerodynamic
  design problems (labelled as instability).

\item These difficulties should have been solved by airframe design
  and not by means of a software patch, i.e. a mechanical solution to
a mechanical problem.
\end{enumerate}
{\sc End promise}
\end{textpromise}

This promise we construct as made by Nader from the following
quote \cite{nader1}:
\begin{quote}
  `And the light at the end of the tunnel is not trying to use a
  hoked-up, glitch-ridden software in the cockpit---MCAS, it's called.
  A software fix for hardware defect? You've got to recall the planes,
  and Boeing has got to develop engineering adjustments and
  engineering changes so that plane is not prone to stall, which is,
  of course, what led to the crashes in Indonesia and Ethiopia,
  killing of 346 innocent people. That's where it's got to be now.'
\end{quote} 

This promise seems to build on the following implicit promise, which we infer (by imputing from the quote):

\begin{textpromise} 
\bigskip\noindent
{\bf Ralph Nader} to {\bf Public}: 

\bigskip\noindent {\sc Promise Body}:
\begin{quote}
  I am knowledgeable about the principles of designing aircraft, and
  about which problems should and should not be solved by means of
  (novel) flight control software.
\end{quote}
{\sc End promise}
\end{textpromise}

\begin{textpromise}(Fundamental solution promise) 

\bigskip\noindent
{\bf Ralph Nader} to {\bf Public}: 

\bigskip\noindent {\sc Promise Body}:
\begin{quote}
  Existing B737 Max planes are best taken back by Boeing for
  adjustment so as to work well without an additional software patch.
\end{quote}
{\sc End promise}
\end{textpromise}
This promise is implicit in the same quote and it seems to presuppose the following promise:

\begin{textpromise}
\bigskip\noindent
{\bf Ralph Nader} to {\bf Public}: 

\bigskip\noindent {\sc Promise Body}:
\begin{quote}
  As an expert in building airplanes I can assure you that it is
  possible to fix the airframe, by way of aerodynamically relevant
  adjustments in such a manner that no additional control software
  component (like MCAS) will be needed.
\end{quote}
{\sc End promise}
\end{textpromise}

How one assesses Nader's comments here, in the form of promises, is of
significance to his credibility as a commentator---just as our
comments above are about us. Assessing them as incorrect could suggest
that Nader's objective was to create reputational damage for Boeing
rather than to offer a safe solution to the flight control issue.
Remarks by law professor W. Bradley Wendel, which we paraphrase from
\cite{Wendel2019} is more accommodating to possible solutions:

\begin{textpromise} 
\bigskip\noindent
{\bf W. Bradley Wendel} to {\bf Public} readers ({\bf Public} in scope)

\bigskip\noindent {\sc Promise Body}:
\begin{quote}
  If one contemplates holding Boeing accountable for the problematic
  design of B737 Max aircraft, then the best approach would be to propose a
  Rational Alternative Design (RAD)---that is a design modification
  which has the following properties: (i) when implemented it prevents
  the problems which have occurred from occurring, (ii) the
  alternative can be proposed by persons with ordinary professional
  knowledge, (iii) the alternative functions properly in all relevant
  circumstances.
\end{quote}
{\sc End promise}
\end{textpromise}

\begin{textpromise} 
\bigskip\noindent
{\bf W. Bradley Wendel} suggests to {\bf Public}:

\bigskip\noindent {\sc Promise Body}:
\begin{quote}
  Rational Alternative Design (RAD) requires that both AOA sensors
  agree when triggering MCAS in an intervention (thereby avoiding
  that the single AOA sensor operates as a
  Single Point of Failure).
\end{quote}
{\sc End promise}
\end{textpromise}
For this matter see \cite{Wendel2019}. 
As a side comment, we mention that~\cite{Wendel2019} takes it for granted
that MCAS primarily provides anti-stall protection (which is in
contrast with promise~\ref{Nas-prom}). That being the case
(hypothetically) a false negative (failing to see that the AOA has
become too high) presents a high risk. Now requiring that both AOA
sensors agree increases the probability of false negatives. This
observation has lead us to promise~\ref{not-agree} above.

Software commentator P. Ladkin takes issue with the assertion of
anti-stall intent, and indicates that he has no conclusive evidence
that software engineers made relevant mistakes, mainly because
spotting system design flaws is not their responsibility
\cite{ladkin1}.

\begin{textpromise}

  \bigskip\noindent {\bf Peter Ladkin} to
  {\bf Public} ({\bf Public} in scope). 

\bigskip\noindent {\sc Promise Body}:
\begin{quote}
  Even if a software update solves the B737 Max problems, there is no
  reason to assume that the original software engineers made mistakes.
\end{quote}
{\sc End promise}
\end{textpromise}

Ladkin points to an issue which is central to the matter: only once it
is known what the software component is supposed to achieve in
principle its engineering can be criticized.

\section{Secondary promises (by Boeing, FAA)}

Boeing issued a communication to the press which indicated
direction in which the problems would be solved.

\begin{textpromise}

\bigskip\noindent
From {\bf Boeing} to {\bf FAA} ({\bf Public} in scope): 

\bigskip\noindent {\sc Promise Body}:
\begin{quote}
  Some improvements and adjustments will be made (leaving the airframe
  unchanged) on the design in order to have it once more certified
  (with in the same type as the B737 NG):
\begin{enumerate}[label=(\roman*)]
\item There will be a warning for the pilots if AOA (angle of attack)
  sensors disagree (beyond 5.5 degrees).

\item Upon detection of a AOA disagreement MCAS-next will be
  deactivated, and at the same time pilots must work with manual
  control.

\item MCAS-next will not repeat its interventions.

\item MCAS-next will take notice of pilot actions and repeated
  attempts to overrule MCAS interventions will be successful.

\end{enumerate}
\end{quote}
{\sc End promise}
\end{textpromise}
The final promise of software learning the pilots' behaviours
throws up new warning flags from a design perspective---namely,
the extent to which {\bf Pilots} can rely on software or hardware
control then becomes completely unknown.

\begin{textpromise}

\bigskip\noindent
From {\bf Boeing} to {\bf FAA} ({\bf Public} in scope): 

\bigskip\noindent {\sc Promise Body}:
\begin{enumerate}[label=(\roman*)]
\item We have proposed adjustments which justify certification before
  the end of 2019.  Prototype airplanes are operational and have made
  many test flights.

  (Notice that the expectation of certification by a deadline is an
  imposition, which is likely without merit.)

\item A high level simulator is available and in this simulator the
improved B737 Max design passes all relevant tests.
\end{enumerate}
{\sc End promise}
\end{textpromise}

\begin{textpromise}
\bigskip\noindent
From {\bf FAA} to {\bf Boeing} ({\bf Public} in scope): 

\bigskip\noindent {\sc Promise Body}:
\begin{quote}
  Certification of the improved B737
  Max will not take place during 2019, and the timing of this action
  is entirely up to the FAA.
\end{quote}
{\sc End promise}
\end{textpromise}
For an outsider comment on the the latter promise we mention~\cite{Cutler2019}.

\section{Secondary promises by the authors}

Let us try to capture part of `what went wrong', using a bundle
of promises. By using promises explicitly, readers may know clearly
our intended meaning, whereas in looser narrative form the points
might be dismissed as rhetoric.
None of what follows should be read as a claim that Boeing engineers
were sloppy or negligent. Their engineering challenge was spectacular,
though perhaps more so in hindsight than at the time of B737 Max
flight control software design. Indeed, it may be fairly difficult to find
out when and where in the development process steps were taken which
could be qualified as defective or problematic.

We assume, in the following, that the B737 Max design is (at least to some extent)
responsible for both accidents, granted any claims that pilot errors,
maintenance problems, and training deficiencies may have been in place
as well. This seems fair, since one must assess the total system and
the fidelity of all agents in keeping their component promises in
promising the outcome of the whole\cite{treatise2}.

\begin{textpromise} (Learning curve risks persist) 

\bigskip\noindent
{\bf Authors} to {\bf Public}. 

\bigskip\noindent {\sc Promise Body}:
\begin{enumerate}[label=(\roman*)]
\item Both catastrophes will in due time be understood as having
  played an important role in the learning curve of aircraft design.
  The idea that deadly catastrophes can be prevented is too
  optimistic.

\item Rather than a focus on which redesign of the B737 Max provides
  the simplest way out of these difficulties, the essential question
  is: in what way are these problems informative about weaknesses of
  the {\bf DO178c} software engineering standards (see
  e.g.~\cite{Pothon2012}).

\item Software engineering may need to take more responsibility of
  that rationale of requirements and specifications. The focus of {\bf
    DO178c} on testing protocols may distract from the promised
  requirements capture and management at higher levels of system
  design. Clearly no amount of classical software testing could have
  prevented the MCAS requirements from being insufficiently flexible
  to deal with that variation of in-flight problems that may occur, without
a meta criterion for finding the failure modes\footnote{We believe that
a promise theoretic methodology could have been of use here, if applies
at the level of system engineering, as mentioned earlier, but this topic
goes beyond the scope of this note.}.
\end{enumerate}
{\sc End promise}
\end{textpromise}

\begin{textpromise}  \label{Maturity}
(Metapromise of B737 maturity.) 

\bigskip\noindent
The {\bf Authors} promise to {\bf Public}:

\bigskip\noindent {\sc Promise Body}:
\begin{quote}
Implicit promises, 
from {\bf Boeing management} and {\bf Staff} were made to {\bf Airlines} and {\bf Pilots}
concerning the B737:
\begin{enumerate}[label=(\roman*)]
\item The B737 NG is a fully mature design with a very good safety track
record. It constitutes the current endpoint of the most successful
development line of commercial jetliners. The B737 constitutes the
pinnacle of airframe reliability.

\item Just as the A320 had to be adapted to new engines, a development
  process which Airbus has carried out in recent years, the B737
  deserved for a new engine option, while using the very successful B737 (NG)
  airframe.

\item All safety systems inside the B737 have been developed into full
  maturity, and have proven to work well in countless circumstances,
  and can therefore be relied upon when adapting the design.
\end{enumerate}
\end{quote}
{\sc End promise}
\end{textpromise}
Note that this latter promise is a promise about another set of
promises.  These have been inferred by us, and the proposed promisers
might not have made these exactly as stated. Promising on behalf of
other agents is a common phenomenon, but it may be flagged in PT as a
violation. Agents cannot make promises on behalf of other agents, as
they are not in possession of their private knowledge or capabilities.
Readers may take this into account in such instances of public discourse.
Some further observations:

\begin{textpromise} (The manual trim wheel as an emerging risk factor.) 

\bigskip\noindent
{\bf Authors} to {\bf Public}:

\bigskip\noindent {\sc Promise Body}:
\begin{enumerate}[label=(\roman*)]
\item Boeing management and engineers have failed to notice that
  during the years of development the trim wheel has become a less
  understood part of the B737 design, mainly because the problem it is
  supposed to solve, viz. a stabilizer runaway, has become so rare due
  to advances in motorized control technology technology for movable
  parts of wings and rudders.

\item Upon the introduction of MCAS, a novel scenario for a problem
  which is similar to but not identical to a classical stabilizer
  runaway, has arisen---but that went somehow unnoticed.

\item It was not recognized, in time, that the stabilizer trim wheel
  for the B737 Max may no longer be the reliable solution
  to stabilizer problems which it used to be in the past. The
  impeccable safety record of parts of the B737 NG could not be
  extended to the new design for this particular part of the B737
  flight control technology, as the agent designs had different capabilities.

\item Even in the absence of MCAS there were grounds to re-evaluate
  the functionality of the manual stabilizer trim wheel: which
  scenario's for a stabilizer runaway exist, and is the manual use of
  the trim wheel a sufficiently reliable option for solving the
  problems in these scenario's.  In other words: for which problem
  scenarios are the trim wheel to be considered a backup option, and
  is it sufficiently usable in those cases? This question must be
  understood in the context of currently fashionable B737 {\bf Pilot}
  training traditions and in the context of {\bf Pilot} practice and
  experience as in existence today.
\end{enumerate}
{\sc End promise}
\end{textpromise}
Many remarks are made on many websites on the  functionality, the use,  and the various
design options for trim control. Obtaining an overview of this topic is not so 
easy (see, for instance \cite{stackex1}). In some airplane designs the FBW (fly by wire)
system takes care of stabilizer trim, except perhaps just before takeoff.
\begin{textpromise} (MCAS functionality: a conceptual problem)

\bigskip\noindent
{\bf Authors} to {\bf Public}:

\bigskip\noindent {\sc Promise Body}:
\begin{quote}
  Explaining what MCAS promised to deliver in the first place is so
  difficult that we have found no single text which explains in any
  detail how that may work?\footnote{%
    In another wording this question appears in a post written by Mark
    R. Jacobsen\cite{RJ1,wiki1} who
    states that MCAS used the stabilizer and changing its position by
    means of its controls to make the pilot feel different forces on
    the yoke. If true the mechanism is novel in civil aviation and the
    FAA should have been much more systematic about its approval.} The
  best approximation we can provide is the following:

\begin{enumerate}[label=(\roman*)]
\item During flight with flaps down and manual control (using the yoke
  buttons for stabilizer trimming) it may be the case that the
  aircraft operates in an angle of attack at which the pilot (having
  the B737 NG in mind) would expect the yoke to require more effort
  to keep in place. By activating the stabilizer trim MCAS the
  creates an episode in which operation of the motor moving the
  jackscrew has as aside effect that the pilot experiences more
  counterforce on the yoke, thereby experiencing what would be felt
  inside a B737 when operating at that same AOA.

\item This side effect makes the pilot experience similar to what would be
experienced in a B737 NG in corresponding conditions. (ii) As a second
effect of MCAS activation the stabilizer is trimmed and the aircraft
is made to turn nose-down, which may be supposed to reduce the angle
of attack more quickly then the pilot would achieved by reacting as
used in a B737: choosing between or combining elevator up, and
additional stabilizer trim.

\item We found no explanation for why MCAS is only supposed to be
active when auto-pilot is off and flaps are down. We assume that, in
auto-pilot mode, there is  no need to make sure that pilot experience is
the same as when operating a B737. In other words: the autopilot must
deal with the same aerodynamic conditions and anomalies, which turns
out to be entirely possible, and for that there is no need to
introduce drastic `external' interventions of stabilizer trimming in
excess of what the auto-pilot as inherited from the B737 NG design
would do.
\end{enumerate}
\end{quote}
{\sc End promise}
\end{textpromise}

\section{The role of transparency}

In a perfect world there would be perfect information---but, in most
cases, we have only partial information. Without perfect information,
logical reasoning is almost impotent, but a reasoning based on
promises can offer plausible outcomes that may be assessed on the
basis of any observer's context. Clearly, the more information one has
about the interior workings of agents, the easier it is to assess
their promised claims. We mention only, only briefly and in passing,
the contemporary tendency to `openness' to public scrutiny, especially
in the software world. The so-called Open Source Software movement is
one example where transparency has been claimed to lead to quality
improvements, by virtue of having `many eyes' look over the details.
There is no such openness in flight systems, so we are reduced to
trusting low resolution promises and making inferences about their
veracity.  The freedom to report (or transparency) of the press, who
report issues, is another case where onlookers have to rely on the
fidelity of the intermediaries who bring us the information. Promise
Theory makes it clear that one has no {\em a priori} reason to trust
relayed information, as it can be distorted intentionally or
unintentionally.  A certain level of openness therefore builds trust.
Unfortunately, neither transparency nor `many eyes' are a guarantee of
truth or even of intent to act in good faith. We know that agents can
be deceived, even with the best of intentions, and that saboteurs can
intentionally seek to deceive.  It would be interesting to study the
role of transparency, and minimum information requirements, in
determining successful models for forensic investigation.  That goes
far beyond our goals here.

\section{Concluding remarks}

The litany of promises above describes a public discussion about the
crash episodes of the Boeing 737 Max. This is to be distinguished from
a technical analysis based on privileged information about the design.
As such, our discussion is more in line with what might be discussed
in a court of law than in a technical fault-finding analysis---it
involves some hearsay and some innuendo, as is the case in all public
assessments.  The relationship between these two viewpoints is not always
acknowledged, but seems to be relevant.

We have laid out the assessable viewpoints of the various agents, with
the sources and recipients labelled clearly. From this, readers can
assess the intentions, each from their own perspectives---this is one
of the values of a Promise Theory approach.  To some extent, such public
discourse inevitably becomes a matter of `he said, she said', and this presents a
challenge to onlookers.  Nevertheless, it is society's responsibility
to mete justice in such cases, so we must take on that challenge.

Promises and impositions also encompass threats as a special case
(see~\cite{Baldwin1971}). In the context of the B737 Max accidents we
have noticed a remarkably prominent occurrence of threats, (quite
apart from the predictable threat that relatives of victims of the
accidents would sue Boeing). Indeed Southwest Airlines expressed its
being unimpressed by the relevance of B737 Max specific simulator
training (see \cite{verge1,bizj1}), and promised to claim 1 million
USD per B737 Max purchased, if specific simulator training would turn
out to be needed.  Boeing accepted this threat and promised the
payments, in the case that it would (after all, and against its
original intentions) prescribe simulator training for the B737
Max\cite{reuter1}. With aircraft getting more complex, providing
extensive simulator training for large numbers of pilots could be a positive
side-effect, and might become a form of system testing, which is likely to become more
relevant and expensive. The promise of automation is normally that it reduces
costs, or improves the human condition. In this case, a half-hearted
attempt at automation seems to have backfired.

From our superficial presentation, constrained by brevity, the benefits
of PT might not be wholly apparent, and may strike the reader as just
another approach to informal logic.  That may be true, but rather than
a focus on evidence and truth PT suggest a focus on the dynamics of
trust and the impact which promising has on the latter. A promise body
may just as well be a lie (see~\cite{Mahon2016}) as it may be a
mathematical theorem.  One may reject our use of the word promise in
which case we refer to~\cite{Gerring1999} for the idea that a
``valid'' PT may allow some deviation of the common understanding of
its keywords in order to arrive at so-called ``good concepts''.  For a
justification of the terminology and its use we refer
to~\cite{BergstraB2014} and the references cited therein, as well as
to~\cite{spacetime1,Burgess2015}.

We have not discussed assessments of promises which have not been
kept.  In an early stage Boeing pointed at pilot errors as explanatory
for the incidents (which is the default strategy for blame in most
accidents), and such remarks may be construed as a promise which has
not been kept and which invites a decrease in many observers' trust in
Boeing.  In a later stage the Boeing CEO stated that Boeing designers
did not miss any details (or fail to spot gaps) in their designs; this
may construed as a promise that is probably not going to be kept.
Benno Baksteen's comments above, rendered as our
promise~\ref{FalseAlarm} are of interest in this regard. Baksteen has
been a (very) public representative of pilots in The Netherlands for
many years. He took sides with Boeing engineers and thereby
unavoidably against the (now dead) pilot teams in his initial
reaction.  This demonstrated a remarkable `reverse default loyalty' to
his constituents. Perhaps this indicates the formidable reputation of
safety for the B737 design line, and reinforces promise~\ref{Maturity}
item (i).

We maintain that MCAS is a proprietary (and closed source) software
component, which implements an algorithm, the MCAS algorithm to which
the {\bf DO178c} standard has been applied to obtain the software
component from a proposed description of the algorithm. To the general
public, this might seem straightforward.  The notion of an algorithm
is somewhat ambiguous, however, and many different definitions have
been given.  We understand an algorithm to be a series of steps for
solving a class of problems, which can be documented as a finite
sequence of instructions. This definition is consistent with
definitions as given in~\cite{CormanLRS2002,CormanLRS2009}, but it
provides some additional detail\footnote{For the notion of an
  instruction sequence and the consequences of requiring finiteness
  thereof we refer to~\cite{BergstraM2012}.}. Without interior details
of the design process, these promises are rather meaningless---which
also renders the notion of design standards somewhat meaningless,
without a formal certificate of compliance. Importantly, in the
current popular use of the word, `algorithm' refers to a software
component in such a manner that the component can be made the subject
of public debate. In other words by speaking of an algorithm one can
make reference to a certain software component without claiming to
know or to understand its technical details. In this way the
ubiquitous use of `algorithm' deviates from its conventional use in
computer science where, in practice, algorithm invariably refers to a
fairly detailed pseudo-code which can be turned in a straightforward
manner into a computer language (e.g. Python), or into whatever
general purpose program notation one may prefer.  The use of the term
algorithm, in the public debate, is more vague however, and
the reader is expected to be aware of that state of affairs.
In the case of aircraft control, and in particular also in the case of
FBW (fly by wire) technology it is common to speak of flight laws as
the underlying concepts from which software components are being
derived. It seems appropriate to use `flight control algorithm' as an
alternative for `flight laws' when discussing software components such
as MCAS in the public domain.

A more thorough approach to mapping out the roles of all involved
agents, their capabilities, and their resulting fidelity in keeping
all their design promises as part of a total system is one possible
future use for Promise Theory's simple methodology. What exists today
seems muddled in checklist semantics without a clear integrated
picture of the system on all scales.  The deliberations, presented
here, surrounding MCAS might also profit from contact with the larger
discussion of Meaningful Human Control (see e.g.~\cite{SantoniH2018}
and~\cite{SlijperBKB2019}), which has become prominent in the theory
of (semi)autonomous systems, including weapons as well as in the
theory and practice of (semi)autonomous driving.

In closing, we note that assessments rooted in personal ideology
cannot easily be avoided in public discourse---not merely by making
use of Promise Theory.  We have adopted the viewpoint that
promise~\ref{Nas-prom}  made from Boeing (that MCAS is not an anti-stall system) is
central to our considerations, and this is crucial.  Were Boeing to
admit that this promise cannot be kept---and therefore was not kept, a
significant loss of trust in Boeing could result, and a consequential
loss of trust in the future of the B737 so large that it might in fact bring
down the whole B737 Max project.  Portraying MCAS as an anti-stall
system by default, as in many contributions to the subject (see, for
example, \cite{adacore}, and in the otherwise outstandingly
informative~\cite{Wendel2019}), runs against the substance of 
promise~\ref{Nas-prom}.  Reference \cite{NicaudBG2001} proposes
speaking of `association' if two notions are identified at some stage,
and may be unidentified at a later stage.  Viewing MCAS as an
anti-stall system may be considered such a temporary association, to
be disassociated if evidence arises that MCAS is not an anti-stall
system. However, there is a key difference between the intent to
behave in a certain way and an emergent and effective behaviour that
seems to do the trick in the moment.  Promise Theoretically we should likely
oppose the validity of emergence, because public safety and legal
assurances are supposed to be based on {\em good intent}.  We consider
promise~\ref{Nas-prom} to be crucial in the analysis of the Boeing 737
Max accidents.

Finally, in~\cite{Sgobba2019} it is argued that the certification
process, conducted by the FAA is in its current form, is unable to
detect some structural risks and that an overhaul of the certification
system is needed in such a manner as to be more focused on risk
analysis.  Without further comment, we add that we do not believe this
to be the case: a checklist style certification process, which merely
enumerates promises at the same level, might well have discovered
a few more relevant weaknesses of the MCAS software component (including software
certification promise lists), but more likely the current
certification methods need a more systematic incorporation into all
levels of engineering---including the relatively new software
development process.

\addcontentsline{toc}{section}{References}
\bibliographystyle{ieeetr}
\bibliography{biblio}

\begin{thebibliography}{10}
\bibitem{stackex1}
Various, ``What are the reasons to evolve from trim tabs to moving the entire
  surface (e.g. stabilizer)?.'' Stack Exchange (Aviation).
\end{thebibliography}

\end{document}